\documentclass[preprint,aps]{revtex4} 

\usepackage{epsf}

\begin{document}

\title{High-Order Coupled Cluster Method (CCM) Calculations Via
  Parallel Processing: An Application To The Kagom\'e Antiferromagnet}

\author{$^*$D.J.J. Farnell, and $^{\dag}$R. F. Bishop}

\affiliation{$^*$Imaging Science and Biomedical Engineering (ISBE), 
  Stopford Building, University of Manchester,
  Oxford Road, Manchester M13 9PT, United Kingdom}

\affiliation{$^{\dag}$
  Department of Physics, University of Manchester Institute of
  Science and Technology (UMIST), P O Box 88, Manchester M60 1QD, United 
  Kingdom}

\date{\today}

\begin{abstract}
A simple ``brute-force'' parallelisation procedure for 
the computational implementation of high-order coupled 
cluster method (CCM) calculations is presented here. This 
approach is investigated and illustrated by an application of 
high-order CCM to the Heisenberg antiferromagnet on the Kagom\'e  
lattice with nearest- and next-nearest-neighbour bonds. 
Ferromagnetic next-nearest-neighbour bonds are used to 
stabilise a model state which contains three sublattices in 
which the spins make angles of 120$^{\circ}$ to each other. 
Ground-state results for up to approximately 10000 fundamental 
clusters are presented, and our best estimate for the 
ground-state energy per spin of the spin-half Kagom\'e 
lattice antiferromagnet with only nearest-neighbour bonds 
is $E_g/N = -0.43104$. We believe that further increases 
(of at least another order of magnitude) in the number of 
fundamental clusters might be possible in future by using 
parallel processing techniques. The extension of high-order 
CCM calculations in order to consider non-N\'eel (e.g., 
dimer solid) model states, simulation of excitation spectra,  
lattice boson and fermion models, and finite-sized systems 
is very briefly considered.
\end{abstract}

\maketitle

\section{Introduction}

The coupled cluster method (CCM) 
\cite{refc1,refc2,refc3,refc4,refc5,refc6,refc7,refc8,refc9}
is a highly successful and widely applied technique of modern-day 
quantum many-body theory. In particular, the CCM has been 
applied to quantum magnetic systems at zero temperature 
\cite{ccm1,ccm2,ccm999,ccm3,ccm4,ccm5,ccm6,ccm7,ccm8,ccm9,ccm10,ccm11,ccm12,ccm13,ccm14,ccm15,ccm16,ccm17,ccm18,ccm19,ccm20,ccm21} 
over the last ten years or so. Indeed, the use of computational 
approaches of the CCM for quantum systems has been 
found to be particularly efficacious. We note that the subsequent 
application of these highly accurate computational
CCM techniques to other types of quantum system
is ripe.

In this article, we present a simple parallelisation procedure 
for high-order coupled cluster method (CCM) calculations. This approach
is investigated and illustrated by the application to the
Heisenberg antiferromagnet on the Kagom\'e lattice with
nearest- and next-nearest-neighbour bonds. In particular,
we note that we may consider up to approximately 10000
fundamental clusters, although we believe that further 
increases (of at least another order of magnitude) in 
the number of fundamental clusters might be possible 
in future by using such techniques.

\section{Formalism}

The exact ket and bra ground-state energy 
eigenvectors, $|\Psi\rangle$ and $\langle\tilde{\Psi}|$, of a 
general many-body system described by a Hamiltonian $H$, 
\begin{equation} 
H |\Psi\rangle = E_g |\Psi\rangle
\;; 
\;\;\;  
\langle\tilde{\Psi}| H = E_g \langle\tilde{\Psi}| 
\;, 
\label{ccm_eq1} 
\end{equation} 
are parametrised within the single-reference CCM as follows:   
\begin{eqnarray} 
|\Psi\rangle = {\rm e}^S |\Phi\rangle \; &;&  
\;\;\; S=\sum_{I \neq 0} {\cal S}_I C_I^{+}  \nonumber \; , \\ 
\langle\tilde{\Psi}| = \langle\Phi| \tilde{S} {\rm e}^{-S} \; &;& 
\;\;\; \tilde{S} =1 + \sum_{I \neq 0} \tilde{{\cal S}}_I C_I^{-} \; .  
\label{ccm_eq2} 
\end{eqnarray} 
The single model or reference state $|\Phi\rangle$ is required to have the 
property of being a cyclic vector with respect to two well-defined Abelian 
subalgebras of {\it multi-configurational} creation operators $\{C_I^{+}\}$ 
and their Hermitian-adjoint destruction counterparts $\{ C_I^{-} \equiv 
(C_I^{+})^\dagger \}$. Thus, $|\Phi\rangle$ plays the role of a vacuum 
state with respect to a suitable set of (mutually commuting) many-body 
creation operators $\{C_I^{+}\}$. Note that $C_I^{-} |\Phi\rangle = 0$, 
$\forall ~ I \neq 0$, and that $C_0^{-} \equiv 1$, the identity operator. 
These operators are furthermore complete in the many-body Hilbert 
(or Fock) space.
Also, the {\it correlation operator} $S$ is decomposed entirely in terms 
of these creation operators $\{C_I^{+}\}$, which, when acting on the 
model state ($\{C_I^{+}|\Phi\rangle \}$), create excitations from it.
We note that although the manifest Hermiticity, 
($\langle \tilde{\Psi}|^\dagger = |\Psi\rangle/\langle\Psi|\Psi\rangle$), 
is lost, the normalisation conditions 
$ \langle \tilde{\Psi} | \Psi\rangle
= \langle \Phi | \Psi\rangle 
= \langle \Phi | \Phi \rangle \equiv 1$ are explicitly 
imposed. The {\it correlation coefficients} $\{ {\cal S}_I, \tilde{{\cal S}}_I \}$ 
are regarded as being independent variables, and the full set 
$\{ {\cal S}_I, \tilde{{\cal S}}_I \}$ thus provides a complete 
description of the ground state. For instance, an arbitrary 
operator $A$ will have a ground-state expectation value given as, 
\begin{equation} 
\bar{A}
\equiv \langle\tilde{\Psi}\vert A \vert\Psi\rangle
=\langle\Phi | \tilde{S} {\rm e}^{-S} A {\rm e}^S | \Phi\rangle
=\bar{A}\left( \{ {\cal S}_I,\tilde{{\cal S}}_I \} \right) 
\; .
\label{ccm_eq6}
\end{equation} 

We note that the exponentiated form of the ground-state CCM 
parametrisation of Eq. (\ref{ccm_eq2}) ensures the correct counting of 
the {\it independent} and excited correlated 
many-body clusters with respect to $|\Phi\rangle$ which are present 
in the exact ground state $|\Psi\rangle$. It also ensures the 
exact incorporation of the Goldstone linked-cluster theorem, 
which itself guarantees the size-extensivity of all relevant 
extensive physical quantities. 

The determination of the correlation coefficients $\{ {\cal S}_I, \tilde{{\cal S}}_I \}$ 
is achieved by taking appropriate projections onto the ground-state 
Schr\"odinger equations of Eq. (\ref{ccm_eq1}). Equivalently, they may be 
determined variationally by requiring the ground-state energy expectation 
functional $\bar{H} ( \{ {\cal S}_I, \tilde{{\cal S}}_I\} )$, defined as in Eq. (\ref{ccm_eq6}), 
to be stationary with respect to variations in each of the (independent) 
variables of the full set. We thereby easily derive the following coupled 
set of equations, 
\begin{eqnarray} 
\delta{\bar{H}} / \delta{\tilde{{\cal S}}_I} =0 & \Rightarrow &   
\langle\Phi|C_I^{-} {\rm e}^{-S} H {\rm e}^S|\Phi\rangle = 0 ,  \;\; I \neq 0 
\;\; ; \label{ccm_eq7} \\ 
\delta{\bar{H}} / \delta{{\cal S}_I} =0 & \Rightarrow & 
\langle\Phi|\tilde{S} {\rm e}^{-S} [H,C_I^{+}] {\rm e}^S|\Phi\rangle 
= 0 , \;\; I \neq 0 \; . \label{ccm_eq8}
\end{eqnarray}  
Equation (\ref{ccm_eq7}) also shows that the ground-state energy at the stationary 
point has the simple form 
\begin{equation} 
E_g = E_g ( \{{\cal S}_I\} ) = \langle\Phi| {\rm e}^{-S} H {\rm e}^S|\Phi\rangle
\;\; . 
\label{ccm_eq9}
\end{equation}  
It is important to realize that this (bi-)variational formulation 
does {\it not} lead to an upper bound for $E_g$ when the summations for 
$S$ and $\tilde{S}$ in Eq. (\ref{ccm_eq2}) are truncated, due to the lack of 
exact Hermiticity when such approximations are made. However, one can prove  
that the important Hellmann-Feynman theorem {\it is} preserved in all 
such approximations. 

In the case of spin-lattice problems of the type considered
here, the operators $C_I^+$ become products of spin-raising
operators $s_k^+$ over a set of sites $\{k\}$, with respect to
a model state $|\Phi\rangle$ in which all spins points 
``downward'' in some suitably chosen local spin axes. 
The CCM formalism is exact in the limit of inclusion of
all possible such multi-spin cluster correlations for 
$S$ and $\tilde S$, although in any real application 
this is usually impossible to achieve. It is therefore 
necessary to utilise various approximation schemes 
within $S$ and $\tilde{S}$. The three most commonly 
employed schemes previously utilised have been: 
(1) the SUB$n$ scheme, in which all correlations 
involving only $n$ or fewer spins are retained, but no
further restriction is made concerning their spatial 
separation on the lattice; (2) the SUB$n$-$m$  
sub-approximation, in which all SUB$n$ correlations 
spanning a range of no more than $m$ adjacent lattice 
sites are retained; and (3) the localised LSUB$m$ scheme, 
in which all multi-spin correlations over all distinct 
locales on the lattice defined by $m$ or fewer contiguous 
sites are retained. 

\section{A ``Brute-Force'' Parallelisation Procedure}

A simple ``brute-force'' approach in order to solve 
the CCM equations in parallel is firstly to rearrange 
Eqs. (\ref{ccm_eq7}) and (\ref{ccm_eq8}) for the ket 
and bra states, respectively, where
\begin{eqnarray} 
{\cal S}_I & = & f_I( ~ {\cal S}_1, ~ \cdots, ~{\cal S}_{I-1}, 
~{\cal S}_{I+1}, ~\cdots,  ~{\cal S}_{N_f} )  \;\; ; \label{ccm_eq10} \\ 
\tilde{{\cal S}}_I & = & \tilde f_I ( ~ {\cal S}_1, ~ \cdots,  ~{\cal S}_{N_f} ;  
~ \tilde {\cal S}_1, ~ \cdots, ~\tilde {\cal S}_{I-1}, 
~\tilde {\cal S}_{I+1}, ~\cdots,  ~\tilde {\cal S}_{N_f} ) 
\;\; ; \label{ccm_eq11} 
\end{eqnarray} 
and where $N_f$ is the number of fundamental configurations. Note 
that there are therefore $N_f$ equations for both $f_I$ and 
$\tilde f_I$, which refer to the ket and bra states, respectively.
We note that each equation contains a finite-number of terms and that,
indeed, this is always the case for the CCM for Hamiltonians 
that contain a finite number of creation and destruction operators.  

The ``brute-force'' approach is to iterate the set 
of equations in Eq. (\ref{ccm_eq10}) to convergence 
first in order to obtain the ket-state correlation 
coefficients. We then iterate the set of equations Eq. 
(\ref{ccm_eq11}) to convergence. This approach appears to 
function adequately as long as one is in the region where
the model state is a good starting point. Clearly, this 
approach will become more difficult as we approach any
{\it critical points} in the CCM equations. 
A simple parallelisation technique is now to split the 
problem of determining and solving the $N_f$ equations 
for both the bra and ket states of Eqs. (\ref{ccm_eq10}) 
and (\ref{ccm_eq11}) between each processor in the 
parallel cluster. Thus, each equation for $I=\{1,\cdots,N_f\}$
is ever dealt with by one node in the cluster only and the 
results of each node are collected together at each 
iteration in the ``brute-force'' algorithm. (Again, we
remember that we iterate the ket-state equations 
first and then the bra-state equations.) Thus, the set 
of equations for the ket and bra states are solved
independently and in parallel.

This approach has a number of advantages. Firstly, we do not 
need to define a Jacobian (e.g., as for the Newton-Raphson 
technique) and so we save on RAM by using this procedure.
Secondly, the task of evaluating and saving the CCM 
equations is reduced by a factor of the number 
of machines used. Indeed, in our implementation
we saved all of the data for a particular set of 
equations to disk locally on each machine, although we
note that one could even re-derive the CCM equations
as and when necessary in order to further 
reduce disk usage. Finally, this a simple
approach and so is easy to amend existing
code and so be parallelised. 

This very simple approach was found to be surprisingly 
successful, although we note that the number of 
iterations for such a direct iteration method is 
at least an order of magnitude greater than that 
for the Newton-Raphson technique, for example.
Indeed, we were able to reproduce those results 
determined on a single machine up to the LSUB7 level 
of approximation by using our parallelised 
implementation. This was found to be an
excellent test of the validity of our results. 
However, we were able to solve the LSUB8 approximation 
for the ground-state energy only by using parallel 
processing. We note that this approximation contained 
10707 fundamental configurations and 
that this is twice as big as the previous 
``largest'' CCM calculation for spin systems. 
We believe that further increases in the number 
of fundamental configurations by at least 
another order of magnitude might be possible 
even by using moderate levels of parallelisation 
(e.g., 10 or 20 nodes) and for our 
``brute-force'' approach.

Finally, we would wish to implement a 
more sophisticated solution of this problem
using parallel processing in the future. A 
possible way of doing this would be to use 
PLAPACK, for example, in order to implement 
the Newton-Raphson technique ``in parallel.''
Other strategies might include numerical approximations
to the Jacobian for the Newton-Raphson technique or 
other less memory-intensive ways of determining 
and solving the CCM equations of Eqs. (\ref{ccm_eq7}) 
and (\ref{ccm_eq8}).

\section{The $J_1$--$J_2$ Kagom\'e Model}

The spin-half $J_1$--$J_2$ Kagom\'e Model model is
is given by
\begin{equation}
H = J_1 \sum_{\langle i , j \rangle} {\bf s}_i \cdot {\bf s}_j
- J_2 \sum_{\{ i , k \}} {\bf s}_i \cdot {\bf s}_k ~~ ,
\label{eq1}
\end{equation}
where $\langle i , j \rangle$ runs over all nearest-neighbour (n.n.)
bonds and $\{ i , k \}$ runs over all next-nearest-neighbour (n.n.n.) 
bonds  on the Kagom\'e lattice. Note that each bond is counted 
once and once only, and that we now set $J_1 = 1.0$.

We choose a model state $|\Phi\rangle$ in which 
the lattice is divided into three sublattices, denoted 
$\{$A,B,C$\}$. The spins on sublattice A 
are oriented along the negative {\em z}-axis, and spins on sublattices 
B and C are oriented at $+120^\circ$ and $-120^\circ$, respectively, 
with respect to the spins on sublattice A. 
Our local axes are chosen by rotating about the 
$y$-axis the spin axes on sublattices B and C by $-120^\circ$ and 
$+120^\circ$ respectively, and by leaving the spin axes on sublattice A 
unchanged. The n.n.n. bonds  for $J_2 > 0$ stabilise 
this type of N\'eel ordering in the quantum ground state.

Careful CCM and finite-sized calculations \cite{ccm18,kaf1,kaf2,kaf3} have 
been performed for the spin-half Kagom\'e antiferromagnet (KAF)
($J_2=0$), and these results indicate that the classical
N\'eel-like ordering observed, for example, in the triangular-lattice
antiferromagnet (TAF) is not seen for the quantum KAF model. The best
estimate for the ground-state energy of the KAF via finite-sized
calculations \cite{kaf3} stands at $E_g/N$=$-$0.43.
Furthermore, CCM and series expansion results \cite{ccm18,kaf2} indicate  
that the ground-state of the KAF is disordered. Indeed,
a variational calculation \cite{kaf4} which utilised a dimerised 
basis also found that the ground state of the KAF 
is some sort of spin liquid. Thus, it is highly 
likely that a quantum phase transition will occur 
at or near to $J_2=0$.

\begin{figure}
\epsfxsize=9cm
\epsffile{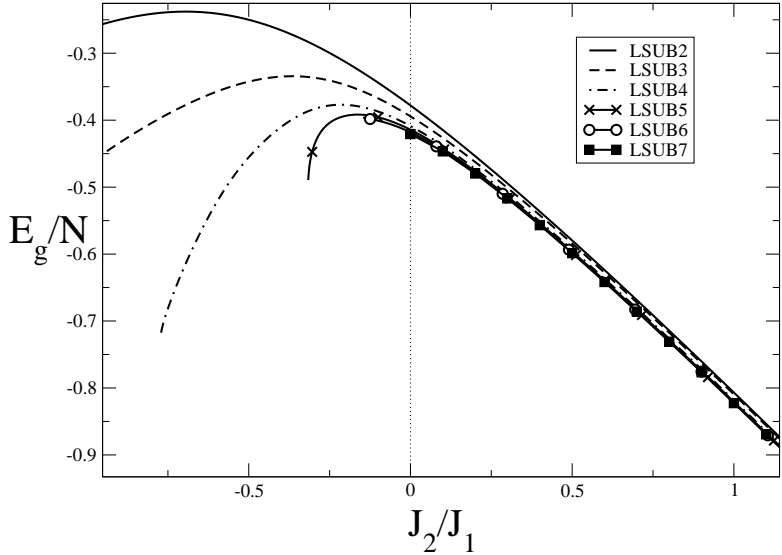}
\vspace{-3.5cm}
\caption{CCM results for the ground-state energy per spin of the 
$J_1$--$J_2$ model (with $J_1=1$) using the LSUB$m$ approximation 
with $m=\{2,3,4,5,6,7\}$. We see that our results converge 
rapidly with $m$ across a wide range of $J_2$ up to an including
the point $J_2=0$.}
\label{fig1}
\end{figure}

Our results for the ground-state energy per spin are given 
in Fig. \ref{fig1} using the LSUB$m$ approximation 
with $m=\{2,3,4,5,6,7\}$. We see that our results converge 
rapidly with $m$ across a wide range of $J_2$ up to an including
the point $J_2=0$. Results for the special case 
$J_2=0$ are given in Table \ref{tab1}. (Note that 
these results are slightly different than previous 
CCM calculations in Ref. \cite{ccm18} due to a different
underlying ``interpolating'' lattice in this case.) We 
are able to consider 10707 fundamental configurations using
our parallel approach and a simple extrapolation in the 
limit $m \rightarrow \infty$ gives $E_g/N = -0.43104$,
which should be considered to previous results of
the best of other calculations \cite{ccm18,kaf3}, 
namely, $E_g/N \approx -0.43$.

\begin{figure}
\epsfxsize=9cm
\epsffile{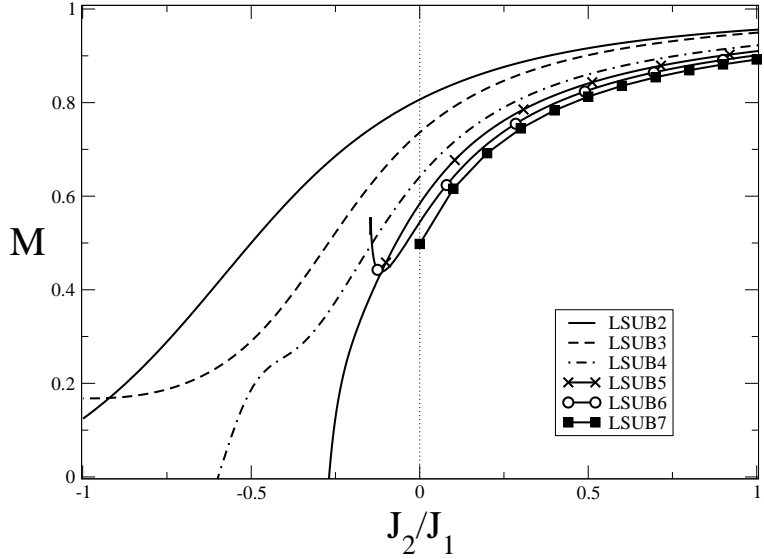}
\vspace{-3.5cm}
\caption{CCM results for the sublattice magnetisation of the 
$J_1$--$J_2$ model (with $J_1=1$) using the LSUB$m$ approximation 
with $m=\{2,3,4,5,6,7\}$. Note that the CCM equations for the LSUB7 
approximation ``break down'' at a critical point near to $J_2=0$, 
and that such behaviour is often associated with a phase transition
in the ``real'' system.}
\label{fig2}
\end{figure}

The sublattice magnetisation is defined by, 
\begin{equation}
M = - \frac 2{N} \sum_{k=1}^{N} s_{k}^z ~~ .
\label{ccm_j_j'_3}
\end{equation}
where $k$ runs over all lattice sites on the Kagom\'e lattice.
Results for the sublattice magnetisation are given 
in Fig. \ref{fig2} using the LSUB$m$ approximation 
with $m=\{2,3,4,5,6,7\}$. Results for the special case 
$J_2=0$ are given in Table \ref{tab1}. Note
that CCM equations break down in the region
$J<0$ at their {\it critical points}. In
the limit $m \rightarrow \infty$ we expect that this
would occur at $J_2=0$. The ``upturn'' in 
$M$ for LSUB6 that occurs in the region $J_2<0$ is an
artifact and such effects occur when the model 
state becomes an increasingly bad starting
point. The CCM equations for LSUB7 break down
at a critical point near to $J_2=0$, and we note 
that our results are consistent with the 
assertion that this model does not demonstrate 
N\'eel ordering at $J_2=0$.

\begin{table}
\caption{CCM results for the ground-state energy per spin and sublattice 
magnetisation of the KAF model ($J_2=0$) using the LSUB$m$ approximation 
with $m=\{2,3,4,5,6,7,8\}$. $N_f$ is the number of fundamental configurations
used. (The symbol ``**'' indicates that this result is yet to be determined.)
Comparison is made in the last row with the results of other calculations. }
\begin{center}
\begin{tabular}{|c|c|c|c|}  \hline 
$m$       &$E_g/N$       &$M$            &$N_f$  \\ \hline\hline
2         &$-$0.377964   &0.806536       &2      \\ \hline
3         &$-$0.394312   &0.736248       &9      \\ \hline
4         &$-$0.408728   &0.641403       &29     \\ \hline
5         &$-$0.414235   &0.583833       &117    \\ \hline
6         &$-$0.418052   &0.544218       &521    \\ \hline
7         &$-$0.420677   &0.497977       &2358   \\ \hline
8         &$-$0.423554   &**             &10707  \\ \hline
{\it c.f.} Refs. \cite{kaf1,kaf2,kaf3,kaf4}
        &$-$0.43
        &0.0
        &--                 \\ \hline
\end{tabular}
\end{center}
\label{tab1}
\end{table}

\section{Conclusion}

We have shown that it is possible to parallelise the
high-order CCM problem implemented using computational
techniques.  A simple ``brute-force'' method was
used, although more sophisticated approaches were
briefly considered.

An application to the spin-half $J_1$--$J_2$ model on the 
Kagom\'e lattice was considered. Results for the 
ground-state energy and sublattice magnetisation
(with respect to three-sublattice N\'eel ordering,
as for the triangular lattice antiferromagnet) was
considered. It was seen that our results were highly
converged over a large range of $J_2$. Our results
were consistent with the assertion that this
model demonstrates no N\'eel ordering at $J_2=0$.
All results were determined using the parallelised
approach and with a single processor, except for a 
single result for the ground-state energy  at $J_2=0$
which contained 10707 fundamental configurations and which
was only able to be carried out ``in parallel.''
Our best estimate for the ground-state energy per spin 
of the spin-half Kagom\'e lattice antiferromagnet
with nearest-neighbour terms only ($J_2=0$)
was $E_g/N = -0.43104$. 

The increasing accuracy of the high-order CCM
technique and its ability to be applied
to even strongly frustrated systems (such as
the model considered here) are strong advantages.
However, there are many possible future extensions
of high-order CCM which would make a parallel 
implementation even more useful.
The CCM has been used previously \cite{ccm10} to determine
the excitation energies of the {\it XXZ} model 
on one-, two-, and three-dimensional lattices 
with great accuracy using ``localised'' approximation
schemes. We note that it is, in principle, 
straightforward to extend these treatments in order
to determine the excitation spectra
of even highly frustrated spin systems 
with equal accuracy and for lattices 
of complex crystallographic symmetries.
Indeed, the use of existing codes would 
make this process simpler.

Such determination of the the excitation
spectra of quantum spin systems using
high-order CCM techniques might be made 
even more useful by the utilisation 
of ``non-N\'eel'' model states. The
model considered here uses just such
a ``N\'eel'' model state and typically 
we perform some kind of notational 
rotation of the spin axis such that
all of the spin ``point downwards''.
This is then the starting point for
our calculations. We note that non-N\'eel
model states (such as the dimer solid
state) have already been used at low
orders of approximation to consider
spin systems which demonstrate 
novel ordering. It would again be 
a straightforward process to develop
our high-order CCM approach for 
the ``localised'' approximation
schemes in order to consider such
non-N\'eel model states.

Another extension of the current calculations
is to bosonic or fermionic systems. Indeed,
the formalism has already been developed
for lattice bosonic systems, which we note
have much in common (from a ``CCM notational''
point-of-view) which large quantum spin
number, $s$, spin systems previously
considered using high-order CCM 
techniques \cite{ccm20}. 

The above calculations were presented for 
the infinite-lattice case, $N \rightarrow
\infty$. We note that it is also possible
to consider the case where $N$ is finite, 
and, again, high-order techniques would be
invaluable here. 

The application of the CCM to quantum spin 
systems at non-zero temperature might 
also be accomplished in future. Indeed,
the application of the CCM at both zero- and 
non-zero temperatures might help to explain
the subtle interplay of quantum and thermal
fluctuations in driving phase transitions
over a wide range of physical parameters.



\begin{thebibliography}{200}


\bibitem{refc1} F. Coester,
  		{\sl Nucl. Phys.} {\bf 7}, 421 (1958);  
		F. Coester and H. K\"ummel, {\em ibid.} {\bf 17}, 477 (1960).

\bibitem{refc2} J. \v{C}i\v{z}ek,
	 	{\sl J. Chem. Phys.} {\bf 45}, 4256 (1966);   
		{\sl Adv. Chem. Phys.} {\bf 14}, 35 (1969).

\bibitem{refc3} R.F. Bishop and K.H. L\"uhrmann,
	 	{\sl Phys. Rev. B} {\bf 17}, 3757 (1978). 

\bibitem{refc4} H. K\"ummel, K.H. L\"uhrmann, and J.G. Zabolitzky, 
		{\sl Phys Rep.} {\bf 36C}, 1 (1978).

\bibitem{refc5} J.S. Arponen, 
		{\sl Ann. Phys.} {\em (N.Y.)} {\bf 151}, 311 (1983).

\bibitem{refc6} R.F. Bishop and H. K\"ummel,
		{\sl Phys. Today} {\bf 40(3)}, 52 (1987).

\bibitem{refc7} J.S. Arponen, R.F. Bishop, and E. Pajanne, 
		{\sl Phys. Rev. A} {\bf 36}, 2519 (1987);
		{\em ibid.} {\bf 36}, 2539 (1987);
		in {\em Condensed Matter Theories}, edited by P. Vashishta,
		R.K. Kalia, and R.F. Bishop (Plenum, New York, 1987), 
		Vol. 2, p. 357.

\bibitem{refc8} R.J. Bartlett,
		{\sl J. Phys. Chem.} {\bf 93}, 1697 (1989).

\bibitem{refc9} R.F. Bishop, 
		{\sl Theor. Chim. Acta} {\bf 80}, 95 (1991). 





\bibitem{ccm1} M. Roger and J.H. Hetherington,  
  {\it Phys. Rev. B} {\bf 41}, 200 (1990); 
  M. Roger and J.H. Hetherington, 
  {\it Europhys. Lett.} {\bf 11}, 255 (1990).

\bibitem{ccm2} R.F. Bishop, J.B. Parkinson, and Y. Xian, 
  {\it Phys. Rev. B} {\bf 44}, 9425 (1991).

\bibitem{ccm999} R.F. Bishop, J.B. Parkinson, and Yang Xian, {\it Phys. 
 Rev. B} {\bf 46}, 880 (1992). 


\bibitem{ccm3} R.F. Bishop, J.B. Parkinson, and Y. Xian, 
  {\it J. Phys.: Condens. Matter} {\bf 5}, 9169 (1993).


\bibitem{ccm4} D.J.J. Farnell and J.B. Parkinson, 
  {\it J. Phys.: Condens.\ Matter}
  {\bf 6}, 5521 (1994).

\bibitem{ccm5} Y. Xian, {\it J. Phys.: Condens.\ Matter} 
  {\bf 6}, 5965 (1994). 


\bibitem{ccm6} R. Bursill, G.A. Gehring, D.J.J. Farnell, J.B. Parkinson, T. 
  Xiang, and C. Zeng, {\it J. Phys.: Condens. Matter} {\bf 7}, 8605 (1995).

\bibitem{ccm7} R. Hale, {\it Ph.D. Thesis}, UMIST Manchester, United
 Kingdom, (1995).

\bibitem{ccm8} R.F. Bishop, R.G. Hale, and Y. Xian,
  {\it Phys. Rev. Lett.} {\bf 73}, 3157 (1994).

\bibitem{ccm9} R.F. Bishop, D.J.J. Farnell, and J.B. Parkinson, 
  {\it J. Phys.: Condens.\ Matter} {\bf 8}, 11153 (1996).

\bibitem{ccm10} D.J.J. Farnell, S.A. Kr\"uger, and J.B. Parkinson, 
  {\it J. Phys.: Condens. Matter} {\bf 9}, 7601 (1997). 

\bibitem{ccm11} R.F. Bishop, Y. Xian, and C. Zeng, in  {\it Condensed 
  Matter Theories}, Vol. {\bf 11}, edited by E.V. Lude\~na, P. Vashishta,
  and R.F. Bishop (Nova Science, Commack, New York, 1996), 
  p. 91.

\bibitem{ccm12} C. Zeng, D.J.J. Farnell, and R.F. Bishop, 
  {\it J. Stat. Phys.}, {\bf 90}, 327 (1998).

\bibitem{ccm13} R. F. Bishop, D.J.J. Farnell, and J.B. Parkinson, 
  {\it Phys. Rev. B}  {\bf 58}, 6394 (1998). 

\bibitem{ccm14} J. Rosenfeld, N.E. Ligterink, and R.F. Bishop,
  {\it Phys. Rev. B} {\bf 60}, 4030 (1999).

\bibitem{ccm15} R. F. Bishop, D. J. J. Farnell, S.E. Kr\"uger, J. B. 
 Parkinson, J. Richter, and C. Zeng, {\it J. Phys.: Condens. Matter}
 {\bf 12}, 7601 (2000).

\bibitem{ccm16}  R.F. Bishop, D.J.J. Farnell, and M.L. Ristig, 
 {\it Int. J. Mod. Phys. B} {\bf 14}, 1517 (2000).

\bibitem{ccm17} S.E. Kr\"uger, J. Richter, J. Schulenberg, D.J.J. Farnell,
 and R.F. Bishop, {\it Phys. Rev. B} {\bf 61}, 14607 (2000). 


\bibitem{ccm18} D.J.J. Farnell, R.F. Bishop, and  K.A. Gernoth,
  {\it Phys. Rev. B} {\bf 63}, 220402R (2001). 


\bibitem{ccm19} D.J.J. Farnell, K.A. Gernoth, and R.F. Bishop,
 {\it Phys. Rev. B}   {\bf 64}, 172409 (2001). 

\bibitem{ccm20} D. J. J. Farnell, K. A. Gernoth, and R. F. Bishop, 
 {\it J. Stat. Phys.} {\bf 108}, 401 (2002).


\bibitem{ccm21} N.B. Ivanov, J. Richter, D.J.J. Farnell, {\it Phys. 
 Rev. B}  {\bf 66}, 014421 (2002).



\bibitem{kaf1} B. Bernu, P. Lechimant, and C. Lhuillier, {\it Physica
  Scripta.} {\bf T49}. 192 (1993).
  
\bibitem{kaf2} P. Lecheminant, B. Bernu, C. Lhuillier, L. Pierre, and 
  P. Sindzingre, {\it Phys. Rev. B} {\bf 56}, 2521 (1997).
  
\bibitem{kaf3} C. Waldtmann, H.-U. Everts, B. Bernu, P. Sindzingre, 
  C. Lhuillier,  P. Lecheminant, and L. Pierre, {\it Eur. Phys. J. B} 
  {\bf 2}, 501 (1998). 
  
\bibitem{kaf4} C. Zeng and V. Elser, {\it Phys. Rev. B} {\bf 51}, 8318
  (1995). 


\end{thebibliography}
\end{document}